\documentclass[12pt]{article}
\begin{document}
\title{Classical solutions of sigma models in curved backgrounds by the Poisson--Lie T-plurality}
\author{Ladislav Hlavat\'y, Jan H\'ybl, Miroslav Turek}
\maketitle
{Faculty of Nuclear Sciences and Physical Engineering, Czech Technical University in Prague, B\v rehov\'a 7, 115
19 Prague 1, Czech Republic}
\\ \\ e-mail: hlavaty@fjfi.cvut.cz, tolavy.vitr@centrum.cz, turekm@km1.fjfi.cvut.cz
\begin{abstract}{Classical equations of motion for three--dimensional $\sigma$--models in curved background are solved by a
transformation that follows from the Poisson--Lie T-plurality and transform them into the equations in the flat
background. Transformations of coordinates that make the metric constant are found and used for solving the flat
model. The Poisson--Lie transformation is explicitly performed by solving the PDE's for auxiliary functions and
finding the relevant transformation of coordinates in the Drinfel'd double. String conditions for the solutions
are preserved by the Poisson--Lie transformations. Therefore we are able to specify the type of $\sigma$--model
solutions that solve also equations of motion of three dimensional relativistic strings in the curved
backgrounds. Simple examples are given}

\end{abstract}


\def\tuc{\bf}
\def\ba{\begin{array}}
\def\ea{\end{array}}
\def\be{\begin{equation}}
\def\ee{\end{equation}}
\def\lbl{\label}
\def \rf  {(\ref}

\def\eqn{equation}
\def\cond{condition}
\def\tfn{transformation}
\def\soln{solution}
\def\fn{function}
\def\sm{$\sigma$--model}
\def\pl{Poisson--Lie}
\def\pltfn{Poisson--Lie transformation}
\def\dd{Drinfel'd double}
\def\vbe{vanishing $\beta$ function equations}
\def\3dial{three--dimensional}
\def\-1{^{-1}}
\def\half{\frac{1}{2}}
\def\coor{coordinate}
\def\real{{\bf R}}
\def\compl{{\bf C}}

\def\e{{\rm e}}
\def\real{{\bf R}}
\def\cd{{\mathcal D}}
\def\cg{{\mathcal G}}
\def\tcg{\widetilde{{\mathcal G}}}
\def\tx{\widetilde{X}}

\def\th{\widetilde{h}}
\def\sm{$\sigma$--model}
\def\pltp{Poisson--Lie T--pluralit}
\def\pltd{Poisson--Lie T--dualit}
\def\dd{Drinfel'd double}

\section{Introduction}
Recently a classical solution of equations of motion of a $\sigma$--model in curved background was solved by the
T--duality transformation \cite{hla:slnbytduality}. Poisson--Lie T-plurality \tfn s provide us with a
prescription how to relate a much wider class of \sm s with apparently different target space geometries. In
this paper we are going to present two cases of \3dial $\sigma$--models that can be solved by this way.
Additional conditions on the solutions then produce classical strings in the curved backgrounds.

Klim\v{c}\'{\i}k and \v{S}evera in their seminal work \cite{klse:dna} described the conditions and procedure for
transforming solutions of a \sm{} to those of a dual one. This procedure can be extended in such a way that it
enables transformation of the solution of original model to solution of every \sm {} that can be derived from
equations on the \dd {} defined by the original \sm.

Let us assume that a covariant second order tensor field $F$ is given on a Lie group $G$. The classical action
of the \sm{} then is \be S_F[\phi]=\int d^2x\,
\partial_- \phi^{\mu}F_{\mu\nu}(\phi)
\partial_+ \phi^\nu \label{sigm1} \ee where the functions $ \phi^\mu:\ \real^2\rightarrow \real,\
\mu=1,2,\ldots,{\dim}\,G$ are obtained by the composition $\phi^\mu=y^\mu\circ\phi \ $ of a map $
\phi:\real^2\rightarrow G$ and a coordinate map $y:U_g\rightarrow \real^n,\ n={\dim}\,G$ of a neighborhood of an
element $\phi(x_+,x_-)=g\in G$. The equations of motion have the form \begin{equation}\label{eqmot}
 \partial_- \partial_+ \phi^{\mu}+\Gamma_{\nu\lambda}^\mu \partial_- \phi^{\nu}\partial_+ \phi^{\lambda}=0
\end{equation}
where
\begin{equation}\label{chri}  \Gamma_{\nu\lambda}^\mu:=\frac{1}{2}G^{\mu\rho}(F_{\rho\lambda,\nu}
+F_{\nu\rho,\lambda}-F_{\nu\lambda,\rho}),
\end{equation}and
$$ G_{\mu\nu}= \half(F_{\mu\nu}+F_{\nu\mu}),\ \ G_{\mu\nu}G^{\nu\rho}=G^{\rho\nu}G_{\nu\mu}=\delta^\rho_\nu.$$

If $F$ satisfies \be {\mathcal L}_{v_i}(F)_{\mu\nu}= F_{\mu\kappa}v^\kappa_j\widetilde
f_i^{jk}v^\lambda_kF_{\lambda\nu}, \ i,\mu,\nu=1,\ldots,\dim\, G \label{klimseveq}\ee where $v_i$ form a basis
of left--invariant fields on $G$ and $\widetilde f_i^{jk}$ are structure coefficients of a Lie group $\widetilde
G, \ \dim\,\widetilde G=\dim\, G$, then one can rewrite the \eqn s of motion of the \sm{} as equations for a map
to a \dd {} -- connected Lie group whose Lie algebra $\cal D$ admits a decomposition
into two subalgebras that are maximally isotropic with respect to a bilinear, symmetric, nondegenerate,
ad--invariant form on $\cd$. This decomposition $\cd =(\cg| \tcg)$ is called the Manin triple.

The fact that several decompositions of its Lie algebra $\cd$ into Manin triples {$(\cg|\tcg$)} may exist for
one \dd {} leads to the notion of Poisson--Lie T--plurality \cite{unge:pltp}. Solutions of the equations of
motion for $S_F$ then can be transformed to \soln s of another \sm {} for $S_{\widehat F}$ where $\widehat
F_{\mu\nu}$ is a second order tensor field on a group $\widehat G$ that belongs to another decomposition of the
same \dd.

In this paper we are going to exploit plural decompositions of two six--dimensional \dd s denoted in the
classification \cite{snohla:ddoubles} as DD11 and DD12. The reason for this choice is that the \pltfn{} relates
\sm s with flat and curved background and meanwhile the \eqn s necessary for performing the \tfn{} are
explicitly solvable.
\section{Elements of \pltfn}
The tensors $F$ of dualizable \sm s, i.e. such that satisfy (\ref{klimseveq}), can be written in terms of
right--invariant fields on a Lie group $G$ as \be F_{\mu\nu}(y)=e_\mu^a (g(y))E_{ab}(g(y))
e_\nu^b(g(y))\label{metorze} \ee where  $y^i$ are local coordinates of $g\in G$, $e_\mu^a$ are components of
right-invariant forms (vielbeins) $e_\mu^a(g)= \left( ({\rm d}g)\cdot g^{-1} \right)_\mu^a$, \be
E(g)=(E_0^{-1}+\Pi(g))^{-1}, \ \ \ \Pi(g)=b(g)a(g)^{-1} = -\Pi(g)^t,\lbl{poiss}\ee and $a(g),b(g)$ are
submatrices of the adjoint representation of the group $G$ on the Lie algebra of the \dd {}
\footnote{ t denotes transposition.} \be Ad(g)^t  =  \left ( \begin{array}{cc}
  a(g)&0  \\ b(g)&d(g)  \end{array} \right ). \lbl{adg}\ee

The Poisson--Lie \tfn{} of the tensor $F$ is defined as follows \cite{unge:pltp}. Let $\{ T_j,\tilde T^k\},\
j,k\in\{1,...,n\}$ {be} generators of Lie subalgebras {$\cg,\,\tcg$} of the Manin triple associated with the
Lagrangian \rf{sigm1}) and $\{ U_j,\tilde U^k\}$ are generators of some other Manin triple {$(\cg_U|\tcg_U)$} of
the same \dd {} related by the $2n\times 2n$ transformation matrix as \be {\left (
\begin{array}{c}
  \vec T \\ \vec {\tilde T} \end{array} \right )}
  =  \left ( \begin{array}{cc}
 P&T  \\ R&S  \end{array} \right )\left ( \begin{array}{c} \vec U\\ \vec {\tilde  U} \end{array} \right),
\lbl{trsfmat}\ee where
$$ \vec T=(T_1, \ldots, T_n)^t, \ldots, \; \vec {\tilde  U} =(\tilde U^1,\ldots,\tilde U^n)^t.$$
The transformed model is then given by {the} Lagrangian of the form \rf{sigm1}) but with $E(g)$ replaced by \be
\widehat E_U(g_u)=M(N+\Pi_U(g_u)\,M)^{-1}=(\widehat E_{0}^{-1}+\Pi_U(g_u)\,)^{-1},\lbl{eg} \ee where
 \be    \widehat E_{0}=MN^{-1},\ \ M=S^tE_0-T^t, \ \
N=P^t-R^t E_0,\ \lbl{mn} \ee
 and $\Pi_U$ is calculated by \rf{poiss})
from the adjoint representation of the group $G_U$ generated by $\{U_j\}$.

The relation between the solutions $\phi(x_+,x_-)$ of the \eqn s of motion of the model given by $F$ and
$\widehat \phi(x_+,x_-)$ of the model given by $\widehat F$ follows from two possible decompositions of elements
$d$ of Drinfel'd double \be d=g.\widetilde h= \widehat g.h' \label{decofdd}\ee where $g\in G,\ \widetilde h\in
\widetilde G,\ \widehat g\in \widehat G,\ h'\in \widetilde{\widehat G}$.

The map $ \widetilde h:\real^2\rightarrow \widetilde G$ that we need for this transformation satisfies the
equations \cite{kli:pltd}

\be ((\partial_+ \widetilde h).\widetilde h\-1)_j = -A_{+,j}:=
-v^\lambda_jF_{\lambda\nu}(\phi)\partial_+\phi^\nu \label{btp}\ee\be ((\partial_- \widetilde h).\widetilde
h\-1)_j = -A_{-,j}:=\partial_-\phi^\lambda F_{\lambda\nu}(\phi)v^\nu_j \label{btm}\ee

The equations (\ref{trsfmat}--\ref{btm}) define the \pl {} transformation but their solution is usually very
complicated to use them for finding the solutions. Essentially, there are three steps in performing the
transformation:
\begin{enumerate}
\item You must know a solution $\phi(x_+,x_-)$ of the \sm{} given by $F$.
\item For the given $\phi(x_+,x_-)$ you must find $\widetilde h(x_+,x_-)$ i.e. solve the system of PDE's (\ref{btp},\ref{btm}).
\item For given $d(x_+,x_-)=\phi(x_+,x_-).\widetilde h(x_+,x_-)\in D$ you must find the decomposition
$d(x_+,x_-)=\widehat \phi(x_+,x_-).h'(x_+,x_-)$ where $\widehat \phi(x_+,x_-)\in \widehat G$, $h'(x_+,x_-)\in
\widetilde {\widehat G}$.
\end{enumerate}
In the next section we shall present two examples of three--dimensional \sm s with nontrivial (i.e. curved)
backgrounds found in \cite{hlasno:3dsm2} for which all the three steps can be done and the \sm s can be
explicitly solved by the \pl {} transformation. The models are obtained from the Manin triples
$(5|1)\cong(6_0|1)$ and $(4|1)\cong(6_0|2)$ (the notation will be explained below) that are decompositions of
the \dd s DD11 and DD12 respectively, and relate \sm s in flat and curved backgrounds.
\section{Solution of the \sm s corresponding to DD11}
The first two \sm s we are going to investigate are given by the tensors
\begin{equation}\label{F51}
F_{\mu\nu}( y)=\left(
                \begin{array}{lll}
                  0 & 0 & \kappa e^{- y_{1}} \\
                  0 & \kappa e^{-2 y_{1}} & 0 \\
                  \kappa e^{- y_{1}} & 0 & 0 \\
                \end{array}
              \right),\ \  \ \kappa \in R,
\end{equation}
and
\begin{equation}\label{F601}
\widehat F_{\mu\nu}(\widehat y)=\left(
 \begin{array}{lll}
    \frac{1}{{\kappa}}e^{-2\widehat y_{3}} & \frac{1}{{\kappa}}e^{-2\widehat y_{3}} & \frac{{\kappa}}{2}e^{\widehat y_{3}} \\
    \frac{1}{{\kappa}}e^{-2\widehat y_{3}} & \frac{1}{{\kappa}}e^{-2\widehat y_{3}} & -\frac{{\kappa}}{2}e^{\widehat y_{3}} \\
    \frac{{\kappa}}{2}e^{\widehat y_{3}} & -\frac{{\kappa}}{2}e^{\widehat y_{3}} & 0 \\
  \end{array}
\right)
\end{equation}
Both these tensors are symmetric so that the models are torsionless and the tensors $F$ and $\widehat F$ can be
interpreted as metrics. The first metric is flat. The Ricci tensor of the second metric is nontrivial so that
the background of the \sm {} given by (\ref{F601}) is curved but its Gauss curvature is zero.

The tensor $F$ can be obtained from the formulas \rf{metorze}--\ref{adg}) using
\begin{eqnarray}\label{E01} E_0 =
\left(%
\begin{array}{ccc}
  0 & 0 & \kappa \\
  0 & \kappa & 0 \\
  \kappa & 0 & 0 \\
\end{array}%
\right)=F|_{y=0},& \kappa \in R,
\end{eqnarray} and Manin triple $(5|1)$ that is a decomposition $\cd =(\cg| \tcg)$ of the \dd{} DD11
where $\cg= Bianchi\,5$ given by the commutation relations
\begin{eqnarray}\lbl{B5}  [T_1,T_2]=-T_2,& [T_2,T_3]=0,&
[T_3,T_1]=T_3,
\end{eqnarray}and $\tcg$ abelian.

The tensor $\widehat F$ is obtained from  $\widehat{E_0} = \widehat{F}|_{\widehat y=0}$ and another
decomposition of DD11, namely $(6_0|1)$ where $\cg= Bianchi\,6_0$ given by the commutation relations
\begin{eqnarray}\lbl{B60} [U_1,U_2]=0 ,& [U_2,U_3]=U_1 ,& [U_3,U_1]=-U_2,
\end{eqnarray} and $\tcg$ abelian.

Relation between these two decompositions is given by \rf{trsfmat}) where
\begin{eqnarray}\label{matice:60151}
\begin{array}{ccc}
\left ( \begin{array}{cc}
 P&T  \\ R&S  \end{array} \right )&=&\left(
\begin{array}{llllll}
0 & 0 & -1 & 0 & 0 & 0  \\
0 & 0 & 0 & 1 & 1 & 0 \\
-1 & 1 & 0 & 0 & 0 & 0 \\
0 & 0 & 0 & 0 & 0 & -1 \\
\frac{1}{2} & \frac{1}{2} & 0 & 0 & 0 & 0 \\
0 & 0 & 0 & -\frac{1}{2} & \frac{1}{2} & 0
\end{array}
\right).
\end{array}
\end{eqnarray}
Inserting (\ref{E01}) and \rf{matice:60151}) into (\ref{mn}) we get $\widehat{E_0} = \widehat{F}|_{\widehat
y=0}$ and when the corresponding matrix \rf{eg}) is inserted into \rf{metorze}) we get the tensor
$\widehat{F}(\widehat y)$. It means that the equations of motion for the \sm s given by both \rf{F51}) and
\rf{F601}) can be rewritten  as the same \eqn{} on the six--dimensional Drinfel'd double DD11. In other words,
the \sm s are \pl{} T--plural. In the following subsections we shall present the \pl{} \tfn {} between solutions
of their \eqn s of motion.
\subsection{Solution of the flat model} Even though we know that the model given by the tensor (\ref{F51}) lives
in the flat background it is not straightforward to solve the equation of motion because the Christoffel symbols
are not zero. To find the functions $\phi^\mu(x_+,x_-)$ that solve the equation of motion we must express
$\phi^\mu$ in terms of coordinates for which the metric become constant. Transformation of coordinates
\cite{tur:dipl}
\begin{eqnarray}\label{transformacexi1}
\begin{array}{c}
\xi_{1}= -e^{-\phi^1}\\ \\
\xi_{2}= \phi^{2}e^{-\phi^1}\\ \\
\xi_{3}=\phi^{3}+\frac{1}{2}(\phi^{2})^{2}e^{-\phi^1}
\end{array}
\end{eqnarray}
brings the metric \rf{F51}) to constant form $ G'(\xi)=E_0$ and equations of motion transform to the wave
equations so that
\begin{equation}\label{solxi2}
    \xi_j(x_+,x_-)=W_j(x_+)+Y_j(x_-)
\end{equation}
with arbitrary $W_j$ and $Y_j$. Functions $\phi^\mu(x_+,x_-)$ that solve the equations of motion for $S_F[\phi]$
then follow from (\ref{transformacexi1}) and (\ref{solxi2}).
\begin{eqnarray}\label{solphi1}
\begin{array}{cl}
{\phi}^1(x_+,x_-)=&-\ln(-W_1-Y_1),\\ \\
{\phi}^2(x_+,x_-)=&-\frac{W_2+Y_2}{W_1+Y_1},\\ \\
{\phi}^3(x_+,x_-)=&W_3+Y_3+\frac{(W_2+Y_2)^2}{2(W_1+Y_1)}.
\end{array}
\end{eqnarray}where $W_j=W_j(x_+),\ Y_j=Y_j(x_-)$.

This finishes the first step, namely obtaining the solution of the flat \sm. The second step of the \pltfn{}
requires solving the system  of  PDEs (\ref{btp},\ref{btm}) for auxiliary \fn s $\widetilde h$.

\subsection{Solution of the system (\ref{btp},\ref{btm})} The coordinates $\widetilde h_\nu$ in the
Abelian group $\widetilde G$ can be chosen  so that the left--hand sides of the equations (\ref{btp},\ref{btm})
are just $\partial_\pm \widetilde h_{\nu}$. The right--hand sides are
\begin{eqnarray}\label{A+} A_+ = \left(
\begin{array}{c}
  \kappa \phi^{3}e^{-\phi^1}\partial_+ \phi^1+\kappa \phi^{2}e^{-2\phi^1}\partial_+ \phi^2+\kappa e^{-\phi^1}\partial_+ \phi^3 \\ \\
  \kappa e^{-2\phi^1}\partial_+ \phi^2  \\ \\
  \kappa e^{-\phi^1}\partial_+ \phi^1
\end{array}%
\right)
\end{eqnarray}
\\
\\
\begin{eqnarray}\label{A-}
A_- = \left(
\begin{array}{c}
  -\kappa \phi^{3}e^{-\phi^1}\partial_- \phi^1-\kappa \phi^{2}e^{-2\phi^1}\partial_- \phi^2-\kappa e^{-\phi^1}\partial_- \phi^3 \\ \\
  -\kappa e^{-2\phi^1}\partial_- \phi^2  \\ \\
  -\kappa e^{-\phi^1}\partial_- \phi^1
\end{array}%
\right)
\end{eqnarray}
and they become rather involved expressions in $W(x_+)$ and $Y(x_-)$ for the solution $\phi^\mu(x_+,x_-)$ found
in the previous subsection. Nevertheless, the equations (\ref{btp},\ref{btm}) can be solved and the general
solution is
\begin{eqnarray}\label{reseni2}
\begin{array}{ccl}
\tilde{h}_1(x_+,x_-)&=&\kappa[Y_{1}(x_{-})W_{3}(x_{+})-Y_{3}(x_{-})W_{1}(x_{+})+\gamma(x_{+})+\delta(x_{-})]\\ \\
\tilde{h}_2(x_+,x_-)&=&\kappa[Y_{1}(x_{-})W_{2}(x_{+})-Y_{2}(x_{-})W_{1}(x_{+})+\alpha(x_{+})+\beta(x_{-})]\\ \\
\tilde{h}_3(x_+,x_-)&=&\kappa[Y_1(x_-)-W_1(x_+)]
\end{array}
\end{eqnarray}
where
\begin{eqnarray}\label{beta}
\alpha\,'&=&W_1W_2'-W_2W_1'\\
\beta\,'&=&Y_2Y_1'-Y_1Y_2'\nonumber
\end{eqnarray}
(primes denote differentiation) and\begin{eqnarray}\lbl{gama}
\gamma\,'&=&W_1W_3'-W_3W_1'\\
\delta\,'&=&Y_3Y_1'-Y_1Y_3'\nonumber
\end{eqnarray}

\subsection{Plural decompositions of elements of the \dd} The final step in the \pl{} transformation follows
from the possibility of rewriting an element of the \dd
$$d(x_+,x_-)=\phi(x_+,x_-).\widetilde h(x_+,x_-)$$ where
$\phi(x_+,x_-)\in G,\ \widetilde h(x_+,x_-)\in \widetilde G,$ to the form $$d(x_+,x_-)=\widehat
\phi(x_+,x_-).h'(x_+,x_-) ,\ \ \widehat \phi(x_+,x_-)\in \widehat G,\ h'(x_+,x_-)\in \widetilde {\widehat G}.$$
As all the goups are solvable we can write all group elements as product of elements of one--parametric
subgroups
\begin{equation}\label{decomp1}
    \e^{\phi^1T_1}\e^{\phi^2T_2}\e^{\phi^3T_3}\e^{\widetilde h_1\widetilde T^1}\e^{\widetilde h_2\widetilde T^2}
    \e^{\widetilde h_3\widetilde T^3}
 = \e^{\widehat \phi^3 U_3}\e^{\widehat \phi^2 U_2}\e^{\widehat \phi^1U_1}
 \e^{h'_3\widetilde U^3}\e^{h'_2\widetilde U^2}\e^{h'_1\widetilde U^1},
\end{equation} where $T_j,\, \widetilde T^k$ are elements of basis $(5|1)$ and $U_j,\, \widetilde U^k$
are elements of basis $(6_0|1)$. Note the reverse order of subgroups on the right--hand side. Inverting the
matrix \rf{matice:60151}) we get from \rf{trsfmat}) the right--hand side in the form $$
e^{\widehat{\phi}^{3}(-T_{1})}e^{\widehat{\phi}^{2}(\frac{1}{2}T_{3}+\widetilde
T^{2})}e^{\widehat{\phi}^{1}(-\frac{1}{2}T_{3}+ \widetilde T^{2})} e^{h'_{3}(-\widetilde T^{1})}e^{h'_{2}
(\frac{1}{2}T_{2}+\widetilde T^{3})}e^{h'_{1}(\frac{1}{2}T_{2}-\widetilde T^{3})}$$ and the two possible
decompositions of the \dd{} elements yield an equation for $\widehat \phi_\mu$ and $h'^\nu$ in terms of
$\widetilde h_\lambda$ and $\phi^\rho$. To solve it might be rather complicated in general but in this case when
the only nonzero Lie products of $(5|1)$ are
\[ [T_1,T_2]= -T_2, \ \ \ [T_1,T_3]= -T_3, \ \ \ [T_1,\widetilde T^2]=\widetilde T^2,\]
\[ [T_1,\widetilde T^3]=\widetilde T^3,\ \ \ [T_2,\widetilde T^2]=-\widetilde T^1,
\ \ [T_3,\widetilde T^3]=-\widetilde T^1, \] it can be done. When we have used the reverse order of subgroups on
the left--hand side of (\ref{decomp1}) then we can use the simple form of the Baker--Campbell--Hausdorff formula
\be \e^A\e^B=\e^B\e^A\e^{[A,B]} \label{zamenag}\ee since in relevant cases $[[A,B],A]=0=[[A,B],B]$. By repeated
application of this formula we get
\begin{eqnarray}\label{hatphi1}
\begin{array}{rl}
\widehat{\phi}^1=&-\phi^3+\frac{1}{2}\tilde{h}_2  \\  \\
\widehat{\phi}^2=&\phi^3+\frac{1}{2}\tilde{h}_2  \\  \\
\widehat{\phi}^3=&-\phi^1   \end{array}
\end{eqnarray}
\begin{eqnarray}
\begin{array}{rl}
h'_1=&-\frac{1}{2}\tilde{h}_3+\phi^2  \\ \\
h'_2=&\frac{1}{2}\tilde{h}_3+\phi^2 \\   \\
h'_3=&-\tilde{h}_1+\tilde{h}_2\phi^2
\end{array}
\end{eqnarray}
Inserting (\ref{solphi1}) and (\ref{reseni2}) into (\ref{hatphi1}) we get the solution of the equations of
motion
for the \sm{} in the curved background 
(\ref{F601})
\begin{eqnarray}\label{solhatphi1}
\begin{array}{cl}
\widehat{\phi}^1(x_+,x_-)=&\frac{1}{2}\kappa\left({Y_1}({x_-}){W_2}({x_+})-{Y_2}({x_-}){W_1}({x_+})\right)
-\left({W_3}({x_+})+{Y_3}({x_-})\right)\\
\\&-\frac{1}{2}{\frac{\left ({W_2}({x_+})+{Y_2}({x_+} )\right
)^{2}}{\left({W_1}({x_+})+{Y_1}({x_-})\right)}}+\frac{1}{2}\kappa(\alpha({x_+})+\beta({x_-})),\\ \\
\widehat{\phi}^2(x_+,x_-)=&\frac{1}{2}\kappa\left({Y_1}({x_-}){W_2}({x_+})-{Y_2}({x_-}){W_1}({x_+})\right)+\left({W_3}({x_+})+{Y_3}({x_-})\right)\\
\\&+\frac{1}{2}{\frac {\left ({W_2}({x_+})+{Y_2}({x_-} )\right
)^{2}}{{W1}({x_+})+{Y_1}({x_-})}}+\frac{1}{2}\kappa (\alpha({x_+})+\beta({x_-})),
\\ \\
\widehat{\phi}^3(x_+,x_-)=&\ln (-{W_1}({x_+})-{Y_1}({x_-})).
\end{array}
\end{eqnarray}where  $\alpha,\ \beta$ satisfy
\rf{beta}). An example of a simple solution is
    \begin{eqnarray}
\begin{array}{cl}
\widehat{\phi}^1(x_+,x_-)=&-\frac{1}{2}\left ({W_1}({x_+})+{Y_1}({x_-} )\right
) \\ \\
\widehat{\phi}^2(x_+,x_-)=&\frac{1}{2}\left ({W_1}({x_+})+{Y_1}({x_-} )\right
)  \\ \\
\widehat{\phi}^3(x_+,x_-)=&\ln(-W_1({x_+})-Y_1({x_-}))\\ \\
\end{array}
    \end{eqnarray}
obtained from (\ref{solhatphi1}) for $W_2=-W_1$, $Y_2=-Y_1$, $W_3=0$, $Y_3=0$, $W_1$ a $Y_1$ arbitrary.
\section{Solution of the \sm s corresponding to DD12}
The \sm s corresponding to the decompositions $(4|1)$ and $(6_0|2)$ of the \dd{} DD12 are given by the tensors
 \be \label{F41}F_{\mu\nu}(y)= \left(%
\begin{array}{lll}
  0 & \kappa\, y_{1}e^{-y_1} & \kappa\, e^{-y_1} \\ \\
  \kappa\, y_{1}e^{-y_1} & \kappa\, e^{-2y_1} & 0 \\ \\
  \kappa\, e^{-y_1} & 0 & 0 \\
\end{array}%
\right),\ \  \kappa\, \in R,
  \ee
and \be \label{F602}\widehat F_{\mu\nu}(\widehat y)=
\left(%
\begin{array}{lll}
  \frac{1}{\kappa\,}e^{2\widehat y_3} & -\frac{1}{\kappa}e^{2\widehat y_3} &
  \frac{1}{2}\kappa\, e^{-\widehat y_3}+e^{\widehat y_3}\widehat y_3
  \\ \\
  -\frac{1}{\kappa\,}e^{2\widehat y_3} & \frac{1}{\kappa\,}e^{2\widehat y_3} & \frac{1}{2}\kappa\,
  e^{-\widehat y_3}-e^{\widehat y_3}\widehat y_3
  \\ \\
  \frac{1}{2}\kappa\, e^{-\widehat y_3}-e^{\widehat y_3}\widehat y_3& \frac{1}{2}\kappa\, e^{-\widehat y_3}
  +e^{\widehat y_3}\widehat y_3& -\kappa\,{\widehat y_{3}}^2
\end{array}%
\right) \ee Both these tensors are torsionless and their symmetric parts can be interpreted as metrics. The
first metric is flat. The Ricci tensor of the second metric is nontrivial but the Gauss curvature is zero.

The tensor $F$ can be obtained from the formulas \rf{metorze}--\ref{adg}) using\begin{eqnarray}\label{E0} E_0 =
\left(%
\begin{array}{ccc}
  0 & 0 & \kappa\, \\
  0 & \kappa\, & 0 \\
  \kappa\, & 0 & 0 \\
\end{array}\right)
\end{eqnarray} and the Manin triple $(4|1)$, i.e.
$\cg=Bianchi\ 4$ given by the commutation relations \begin{eqnarray}\lbl{B4}  [T_1,T_2]=T_3-T_2,& [T_2,T_3]=0,&
[T_3,T_1]=T_3,
\end{eqnarray}and $\tcg$ abelian.

The tensor $\widehat F$ is obtained from  $\widehat{E_0} = \widehat{F}|_{\widehat y=0}$ and the Manin triple
$(6_0|2)$, where $\cg=Bianchi\ 6_0$ is given by the commutation relations
\begin{eqnarray}\lbl{B60b} [U_1,U_2]=0 ,& [U_2,U_3]=U_1 ,& [U_3,U_1]=-U_2,
\end{eqnarray}and
$\tcg= $ Bianchi2 given by the commutation relations
\begin{eqnarray}\label{komutace2}
[\widetilde{U}^1,\widetilde{U}^2]=\widetilde{U}^3 ,& [\widetilde{U}^2,\widetilde{U}^3]=0 ,&
[\widetilde{U}^3,\widetilde{U}^1]=0.
\end{eqnarray}Relation between these two decompositions
is given by \rf{trsfmat}) where
\begin{eqnarray}\label{matice:60241}
\begin{array}{ccc}
\left ( \begin{array}{cc}
 P&T  \\ R&S  \end{array} \right )&=&\left(
\begin{array}{llllll}
0 & 0 & 1 & 0 & 0 & 0  \\
0 & 0 & 0 & 1 & -1 & 0 \\
1 & 1 & 0 & 0 & 0 & 0 \\
0 & 0 & 0 & 0 & 0 & 1 \\
\frac{1}{2} & -\frac{1}{2} & 0 & 0 & 0 & 0 \\
0 & 0 & 0 & \frac{1}{2} & \frac{1}{2} & 0
\end{array}
\right)
\end{array}
\end{eqnarray}
Inserting (\ref{E0}) and \rf{matice:60241}) into (\ref{mn}) we get $\widehat{E_0} = \widehat{F}|_{\widehat y=0}$
and when the  matrix $E_U(g_u)$ is inserted into \rf{metorze}) the tensor $\widehat{F}$ is obtained. It means
that the equations of motion for the \sm s given by both \rf{F41}) and \rf{F602}) can be rewritten  as the same
\eqn{} on the Drinfel'd double DD12. In the following subsections we shall present the \pl{} \tfn {} between \sm
s given by $F$ and $\widehat F$.
\subsection{Solution of the flat model} To find the functions $\phi^\mu(x_+,x_-)$ that solve the equation
of motion of the flat \sm{} we must express $\phi^\mu$ in terms of coordinates $\xi$
\begin{eqnarray}\label{transformacexi}
\begin{array}{c}
\xi_{1}= -e^{-\phi^1}\\ \\
\xi_{2}= e^{-\phi^1}(1+\phi^{2})+\phi^1-1\\ \\
\xi_{3}=\frac{1}{2}e^{-\phi^1}(1+\phi^2)^2+\phi^3+\phi^1\phi^2-\phi^2 +\phi^1-\frac{1}{2}e^{\phi^1}
\end{array}
\end{eqnarray}for which the metric become
constant and equations of motion transform to the wave equations.
Solution of the equations of motion for $S_F[\phi]$ that follow from (\ref{transformacexi}) and (\ref{solxi2})
then is
\begin{eqnarray}\label{solphi2}
\begin{array}{cl}
{\phi}^1(x_+,x_-)=&-\ln(-W_1-Y_1),\\ \\
{\phi}^2(x_+,x_-)=&-1-\frac{1+W_2+Y_2+\ln(-W_1-Y_1)}{W_1+Y_1},\\ \\
{\phi}^3(x_+,x_-)=&-1+W_3+Y_3+\frac{1}{2(W_1+Y_1)}\left[-1+(W_2+Y_2)^2-(1+\ln(-W_1-Y_1))^2\right].
\end{array}
\end{eqnarray}where $W_j=W_j(x_+),\ Y_j=Y_j(x_-)$ are arbitrary \fn s.

As the next step we have to solve the \eqn s for the auxiliary \fn s $\widetilde h_\nu(x_+,x_-)$.
\subsection{Solution of the
system (\ref{btp},\ref{btm})} The coordinates $\widetilde h_\nu$ in the Abelian group $\widetilde G$ can be
chosen  so that the left--hand sides of the equations (\ref{btp},\ref{btm}) are just $\partial_\pm \widetilde
h_{\nu}$. The right--hand sides in this case are
\begin{eqnarray}\label{A+b} A_+ = \left(
\begin{array}{c}
  \kappa\, e^{-\phi^1}( (\phi^2\phi^1- \phi^2+\phi^3)\partial_+ \phi^1+(\phi^1+e^{-\phi^1}\phi^2)\partial_+ \phi^2+
  \partial_+ \phi^3) \\ \\
  \kappa\, e^{-\phi^1}(\phi^1\partial_+ \phi^1+e^{-\phi^1}\partial_+ \phi^2)  \\ \\
  \kappa\, e^{-\phi^1}\partial_+ \phi^1
\end{array}%
\right)
\end{eqnarray}
\\
\\
\begin{eqnarray}\label{A-b}
A_- = \left(
\begin{array}{c}
  -\kappa\, e^{-\phi^1}( (\phi^2\phi^1- \phi^2+\phi^3)\partial_- \phi^1+(\phi^1+e^{-\phi^1}\phi^2)\partial_- \phi^2+
  \partial_- \phi^3) \\ \\
  -\kappa\, e^{-\phi^1}(\phi^1\partial_- \phi^1+e^{-\phi^1}\partial_- \phi^2)  \\ \\
  -\kappa\, e^{-\phi^1}\partial_- \phi^1
\end{array}%
\right)
\end{eqnarray}
where the \fn s $\phi^\mu(x_+,x_-)$ are given by \rf{solphi2}). The general solution of the equations
(\ref{btp},\ref{btm}) is
\begin{eqnarray}\label{reseni1}
\begin{array}{ccl}
\tilde{h}_1&=&\kappa\,[Y_2(x_-)-W_2(x_+)+W_1(x_+)(Y_2(x_-)-Y_3(x_-))\\ \\
& &-Y_1(x_-)(W_2(x_+)-W_3(x_+))-\alpha(x_+)-\beta(x_-)+\gamma(x_+)+\delta(x_-)]\\ \\
\tilde{h}_2&=&\kappa\,\left[W_2(x_+)Y_1(x_-)-W_1(x_+)Y_2(x_-)+\alpha(x_+)+\beta(x_-)\right] \\ \\
\tilde{h}_3&=&\kappa\,[Y_1(x_-)-W_1(x_+)]
\end{array}
\end{eqnarray}
where the \fn s $\alpha,\,\beta,\,\gamma,\,\delta$ satisfy \rf{beta}), \rf{gama}).
\subsection{Plural decomposition of elements of the \dd} The possibility of writing an element of the \dd{} in the
decompositions $(4|1)$ and $(6_0|2)$
\begin{equation}\label{decomp2}
    \e^{\phi^1T_1}\e^{\phi^2T_2}\e^{\phi^3T_3}\e^{\widetilde h_1\widetilde T^1}\e^{\widetilde h_2\widetilde T^2}
    \e^{\widetilde h_3\widetilde T^3}
 =  \e^{\widehat \phi^3 U_3}\e^{\widehat \phi^2 U_2}\e^{\widehat \phi^1U_1}
 \e^{h'_3\widetilde U^3}\e^{h'_2\widetilde U^2}\e^{h'_1\widetilde U^1}.
\end{equation} yield an equation for $\widehat \phi_\mu$ and $h'^\nu$ in terms of $\widetilde h_\lambda$ and
$\phi^\rho$. Inverting the matrix \rf{matice:60241}) we get from \rf{trsfmat}) the right--hand side in the form
$$ e^{\widehat{\phi}^{3}T_{1}}e^{\widehat{\phi}^{2}(\frac{1}{2}T_{3}-\widetilde
T^{2})}e^{\widehat{\phi}^{1}(\frac{1}{2}T_{3}+ \widetilde T^{2})} e^{h_{3}'\widetilde T^{1}}e^{h_{2}'
(-\frac{1}{2}T_{2}+\widetilde T^{3})}e^{h_{1}'(\frac{1}{2}T_{2}+\widetilde T^{3})}.$$  The only nonzero Lie
products of $(4|1)$ are
\[ [T_1,T_2]= T_3-T_2, \ \ \ [T_3,T_1]= T_3,\]
\be [T_1,\widetilde T^2]=\widetilde T^2,\ \  \ [T_1,\widetilde T^3]=-\widetilde T^2+\widetilde T^3,\ee
 \[ [T_2,\widetilde T^2]=-\widetilde T^1, \ \  [T_2,\widetilde T^3]=\widetilde T^1, \ \  [T_3,\widetilde T^3]=-\widetilde
 T^1 \]
and when we have used the reverse order of subgroups on the left--hand side of (\ref{decomp1}) only the simple
form of the Baker--Campbell--Hausdorff formula \rf{zamenag}) is necessary in relevant cases.  We get
\begin{eqnarray}\label{hatphi2}
\begin{array}{rl}
\widehat{\phi}^1=&\phi^3+\frac{1}{2}\tilde{h}_2    \\ \\
\widehat{\phi}^2=&\phi^3-\frac{1}{2}\tilde{h}_2    \\ \\
\widehat{\phi}^3=&\phi^1   \end{array}
\end{eqnarray}
\begin{eqnarray}
\begin{array}{rl}
h_1'=&\frac{1}{2}\tilde{h}_3+\phi^2   \\ \\
h_2'=&\frac{1}{2}\tilde{h}_3-\phi^2   \\ \\
h_3'=&\tilde{h}_1-\tilde{h}_2\phi^2+\frac{1}{2}\phi^2\phi^2+\frac{1}{2}\phi^2\tilde{h}_3-\frac{1}{8}\tilde{h}_3\tilde{h}_3                           \\ \\
\end{array}
\end{eqnarray}
Inserting (\ref{solphi2}) and (\ref{reseni1}) into (\ref{hatphi2}) we get the solution of the equations of
motion for the \sm{} given by the action $S_{\widehat F}$ for $\widehat F$ given by (\ref{F602}).
\begin{eqnarray}\label{soltilphi2}
\begin{array}{cl}
\widehat{\phi}^1(x_+,x_-)=&-1+(W_3+Y_3)+\frac{1}{2}\kappa\,\left[W_2Y_1-W_1Y_2+\alpha+\beta\right]+\\
\\& \frac{1}{2(W_1+Y_1)}\left[-2+(W_2+Y_2)^2-2\,\ln(-W_1-Y_1)-\ln^2(-W_1-Y_1)\right]\\ \\
\widehat{\phi}^2(x_+,x_-)=&-1+(W_3+Y_3)-\frac{1}{2}\kappa\,\left[W_2Y_1-W_1Y_2+\alpha+\beta\right]+\\
\\& \frac{1}{2(W_1+Y_1)}\left[-2+(W_2+Y_2)^2-2\,\ln(-W_1-Y_1)-\ln^2(-W_1-Y_1)\right]\\ \\
\widehat{\phi}^3(x_+,x_-)=&-\ln(-W_1-Y_1)    \\ \\
\end{array}
\end{eqnarray}
where $W_j=W_j(x_+),\ Y_j=Y_j(x_-),\ \alpha= \alpha(x_+),\ \beta=\beta(x_-)$ and $\alpha,\ \beta$ satisfy
\rf{beta}). A simple solution dependent on both $x_+$ and $x_-$ is
    \begin{eqnarray}
\begin{array}{cl}
\widehat{\phi}^1=&-\frac{1}{2}(W_2+Y_2)^2+Y_2 \\ \\
\widehat{\phi}^2=&-\frac{1}{2}(W_2+Y_2)^2+W_2\\ \\
\widehat{\phi}^3=&0\\ \\
\end{array}
    \end{eqnarray}
obtained from (\ref{soltilphi2}) for $\kappa = 1$, $W_1=0$, $Y_1=-1$, $W_3=\frac{W_2}{2}$, $Y_3=\frac{Y_2}{2}$,
$W_2$ a $Y_2$  arbitrary.

\section{String solutions}\lbl{stringcond}
It is well known that the string action can be replaced by the \sm{} (Polyakov) action \rf{sigm1}) if an
additional constraint for the \sm{} \soln s \be \partial_a\phi^\mu
G_{\mu\nu}(\phi)\partial_b\phi^\nu=\eta_{ab}e^\omega \ee is required where $$\eta=\left( \begin{array}{cc}
 0&1  \\ 1&0  \end{array} \right )$$ 
and $\omega$ is a function of $x_+,x_-$. It is easy to see that for the flat models given by (\ref{F51}) or
(\ref{F41}) this condition yields \be\lbl{sc1} 2\,W_1'(x_+)\,W_3'(x_+)+W_2'(x_+)\,W_2'(x_+)=0,\ee \be\lbl{sc2}
2\,Y_1'(x_-)\,Y_3'(x_-)+Y_2'(x_-)\,Y_2'(x_-)=0. \ee
We have checked that these conditions 
remain preserved for the curved models as well so that the string solutions both in the flat and curved
backgrounds
can be obtained by inserting a solution of \rf{sc1}), (\ref{sc2}) into \rf{solphi1}), \rf{solhatphi1}),
\rf{solphi2}) and \rf{soltilphi2}).

An example of solution of \rf{sc1}), (\ref{sc2}) is
\begin{eqnarray}\label{string1}
    W_2(x_+)=k_1 W_1(x_+)+k_0,\ \  Y_2(x_-)=c_1 Y_1(x_-)+c_0,\nonumber\\
   W_3(x_+)=-\half k_1^2 W_1(x_+)+k_0,\ \ Y_3(x_-)=-\half c_1^2 Y_1(x_-)+c_0, \end{eqnarray} giving \be
    \alpha(x_+)=-k_0 W_1(x_+),\ \ \beta(x_-)= c_0 Y_1(x_-)+c_2,\ee
where $k_j,\ c_j$ are constants and $W_1,Y_1$ arbitrary functions. Other solutions are three--dimensional open
strings in the light--cone gauge $$\xi_1(x_+,x_-)= W_1(x_+)+Y_1(x_-)={2}\,k\,p^+ \tau=\sqrt{2}\,k\,p^+ (x_+ +
x_-)$$ i.e.
\be W_1(x)=Y_1(x)=\sqrt{2}\,k\,p^+ x \ee 
\be W_2(x)=Y_2(x)=\half X_2+
\sqrt{k}\,\alpha_0\,x+ i\sqrt{\frac{k}{2}}\sum_{n\neq 0}\frac{1}{n}\alpha_n \e^{-in\sqrt{2} x} \ee 
\be W_3(x)=Y_3(x)=\half X^--\frac{i}{2p^+}\sum_{n\neq 0}\frac{1}{n} L_n \e^{-in\sqrt{2} x} \ee and \be
\alpha(x)=-\beta(x)-C={k\,p^+}\left(-\frac{X_2}{\sqrt{2}}\,x+\sqrt{k}\sum_{n\neq 0}\alpha_n
(\frac{i\,x}{n}+\frac{\sqrt{2}}{n^2})\,\e^{-in\sqrt{2} x}\right) , \ee where \be L_n=\half\sum_{p\in{\bf
Z}}\alpha_p\alpha_{n-p},\ee and $k,p^+,X_2,X^-,\alpha_n,C$ are constants.
\section{Conclusions} \pltfn{} between classical solutions of \sm s can be explicitly performed for
the \dd s with sufficiently simple algebraic structure. We have used this fact in solution of equations of
motion of three--dimensional \sm s in the curved backgrounds \rf{F601}) and \rf{F602}). These models are by
construction dual to flat ones.

We have transformed group coordinates of the dual flat models to those for which their metrics are constant so
that solution of the flat \sm{} in the latter coordinates reduces to the solution of the wave \eqn. Afterwards
we had to solve the \eqn s for the auxiliary \fn s necessary for rewriting the \eqn s for the \sm{} to \eqn s on
the \dd{}. Because one of the subgroups of the decompositions of the \dd{} was Abelian, the system of PDEs
(\ref{btp}) and (\ref{btm}) separated and became solvable. Performability of the last step of the \pl{}
transformation, namely finding coordinates of the plural decompositions, depends critically on the complexity of
the structure of \dd{} where the \sm s live. In the investigated cases the Lie algebraic was simple enough so
that the formula (\ref{zamenag}) for solution of (\ref{decomp1}) could be used and coordinates of the plural
decompositions were found.

In the end we have investigated the conditions the must be applied to the \sm{} \soln{} in order that they
satisfy the equations of motion for the relativistic strings. The conditions have very simple form for the flat
models and they are preserved by the \pltfn s. Therefore we were able to specify the type of \sm{} \soln s that
are also \soln s of \eqn s of motion of three dimensional relativistic strings in the curved backgrounds.
Examples were given in the section \ref{stringcond}.
\section{Acknowledgements}
This work was supported by the project of the Grant Agency of the Czech Republic No. 202/06/1480 and by the
research plan LC527 15397/2005--31 of the Ministry of Education of the Czech Republic. The authors are grateful
to Libor \v Snobl for valuable comments.

\end{document}